\documentstyle[12pt,epsfig]{article}
\newcommand{\bm}[1]{\mbox{\boldmath $#1$}}
\def\be{\begin{equation}}
\def\ee{\end{equation}}
\title{Interference phenomena in scalar transport induced by a noise 
finite correlation time}
\author{A.~Mazzino$^{1,2}$ and P.~Castiglione$^{3}$\\
\small{$^{1}$ CNRS, Observatoire de Nice, B.P. 4229,
06304 Nice Cedex 4, France.}\\
\small{$^{2}$ INFM -
Dipartimento di Fisica, Universit\`a di Genova, I--16146 
Genova, Italy.}\\
\small$^3$ INFM -
Dipartimento di Fisica, Universit\`a ``La Sapienza'', I--00185 
Roma, Italy.}
\begin{document}
\maketitle
\date{}
\vspace*{-0.6cm}
\begin{abstract}
The role played on the scalar transport by a finite, not small, 
correlation time, $\tau_u$, for the noise velocity 
is investigated, both analytically and numerically. 
For small $\tau_u$'s a mechanism leading to enhancement of transport 
has recently been identified and shown to be dominating 
for any type of flow. For finite non-vanishing $\tau_u$'s we recognize
the existence of a further mechanism associated with 
regions of anticorrelation of the Lagrangian advecting velocity.
Depending on the extension of the anticorrelated regions, either an 
enhancement (corresponding to constructive interference) or a depletion 
(corresponding to destructive interference) in the turbulent transport
now takes place.
\end{abstract}


\vspace{4mm}
The understanding of the mechanisms  leading to transport 
enhancement/depletion is of great interest in various domains
of science \cite{Moffatt}. In the theory of scalar transport,
the main regime of interest corresponds to 
the long-time behavior of
the concentration field (e.g.~the density of passive particles injected into 
the  flow) for time scales much longer than the characteristic 
time-scale of the advecting velocity field. In this case,
long-wave disturbances governs the slow modes of concentration 
and the asymptotic dynamics is described by a diffusive equation with 
renormalized  diffusion coefficients, $D_{\alpha\beta}^{E}$, namely:
\begin{equation}
\partial_t \langle\theta\rangle= D^{E}_{\alpha\beta}\,
             {\partial^2\over\partial x_\alpha\, \partial x_\beta}\,
               \langle\theta\rangle
\label{eq:1.4}
\end{equation}
where $\langle\theta\rangle$ is the concentration field 
averaged locally over a
volume of linear dimensions much larger than the typical length $l_0$ of the
advecting (incompressible) velocity field.
The latter equation can be rigorously proved by using, for instance, 
asymptotic methods like multiscale techniques \cite{BLP78}.
 From the Lagrangian viewpoint, the large-scale diffusion regime
is  a consequence of the central limit theorem, stating  (roughly)
that, for particle displacements weakly correlated,
their sum tends to a Gaussian and, as an immediate consequence,
particles undergo diffusion. 


An exact (i.e.~non-perturbative) expression of the Green-Kubo type 
obtained by Taylor \cite{T21} allows one
to express the effective diffusivities as time-integrals of
Lagrangian correlations. For Lagrangian chaotic 
flow the correlations decay rapidly, the integral converges 
and gives a {\em finite }  effective diffusivity $D^{E}$.

 In many geophysical domains (for instance, 
to study the dispersion processes either 
in atmospheric or oceanic flows), Lagrangian trajectories, $\bm{x}(t)$, of  
particles advected by a prescribed
velocity field, $\bm{U}(\bm{x}(t))$, and plunged in a noise, small-scale 
velocity, $\bm{u}(t)$, are properly modelled by the Langevin equation:
\begin{equation}
\frac{d\bm{x}(t)}{dt}=\bm{U}(\bm{x}(t)) + \bm{u}(t) 
\label{lange}
\end{equation}
where the small-scale turbulence, $\bm{u}(t)$, is  
described by a Markovian process with finite correlation time, 
$\tau_u$, rather than by a white-noise process, on account of experimental
observations \cite{D85,PN89}
suggesting the presence of some form of finite correlation
for the smallest scales of motion.\\
In the following, the advecting velocity field $\bm{U}(\bm{x}(t))$ will be 
assumed incompressible, the noise term $\bm{u}(t)$ a Gaussian random process
having zero average value and the colored-noise correlation function given by:
\begin{equation}
\langle u_\alpha ( t ) \; u_\beta (t') \rangle = 
\frac{D^e}{\tau_u} \; \delta_{\alpha\beta}
 e^{- \frac{ \mid t-t' \mid }{\tau_u}}\;\;\; ,  
\label{cnoise}
\end{equation}
where $D^e$ is the isotropic eddy-diffusivity arising from the smallest 
scales of turbulent motion 
(not explicitly solved) and $\tau_u$ is their correlation time.

The main question addressed in the present Letter, concerns the role played 
by the finite (not necessarily small) correlation time,  $\tau_u$, 
of the noise velocity in the large-scale dynamics and thus in the  
behavior of the effective diffusivity,  $D^E$ . 
Specifically, we are interested in the difference
\begin{equation}
I(D^e,\tau_u) \equiv D^E(D^e,\tau_u)- D^E(0,0) - D^e
\label{interfer}
\end{equation}
for small, fixed, $D^e$ and different values of $\tau_u$.
The argument of $D^E$
is meant to stress that the eddy-diffusivity depends on both $D^e$ and $\tau_u$
due to the dependence of Lagrangian
trajectories on the colored-noise, as a whole.\\
Physically, positive (negative) values of $I$ correspond 
to constructive (destructive) interference induced by the dependence  
of  Lagrangian trajectories on the colored-noise.\\
In Refs.~\cite{Saff} and \cite{MV97} the interference problem was
investigated focusing the attention on the variation of the sign 
of $I$ for small $D^e$ and $\tau_u=0$ (i.e.~for white-in-time processes).
It was conjectured in Ref.~\cite{Saff}
that the interference process 
should always be destructive, and actually shown in Ref.~\cite{MV97}
that it can be either constructive or destructive, depending on the extension
of the anticorrelated regions of the velocity field.\\
The effect of a small $\tau_u$ on the variation of $I$
was investigated in Ref.~\cite{CM98}. 
Accordingly, splitting $I$ into two distinct parts,
$I=I_{D^e}+I_{\tau_u}$, the first one being the interference contribution
of $D^e$ when $\tau_{u}=0$, and the second the correction due to a
non-vanishing $\tau_{u}$, it was shown perturbatively in the 
parameter $\tau_{u}/\tau_{\mbox{\tiny U}}\ll 1$, 
that $I_{\tau_u}>0$ for any type of flow,
$\tau_{\mbox{\tiny U}}$  being the correlation time of the advecting
component 
${\bm U}$. Physically, the mechanism leading to positive interference,
and thus to enhanced diffusion, was identified in the augmentation
of coherence in the diffusion process induced by the small correlation
time $\tau_{u}$. Indeed, the latter
makes the particles of the diffusing 
substance to forget less rapidly their past 
than in the presence of white-noise turbulence. 
This fact increases the Lagrangian correlation time and therefore the 
eddy-diffusivity.

In this Letter, we want to remark that the above mechanism is certainly 
present for all values
of $\tau_u$, but when the latter is large enough another effect 
can be even stronger.  To show this, let us 
consider the case when the Lagrangian correlation function
presents negative correlated regions.  The contribution given  
by such regions in the integral along the trajectories 
required in order to obtain the eddy-diffusivities 
is negative and a  reduction of  the diffusion may occur.
The role played by $\tau_u$ is thus twofold. On one hand, according to 
the results of Ref.~\cite{CM98}, 
it permits the particles to go away from their initial positions
more slowly. This effect increases the Lagrangian correlation 
time and therefore $D^E$.
On the other hand, it works to maintain the particles inside the trapping
zones where the velocity is anticorrelated. 
The slowing-down due to these anticorrelated regions is thus enhanced and 
diffusion is reduced. Due to the well-known
sensitivity of transport to slow parts of the trajectories,
the latter mechanism can become stronger than the former and 
a reduction in the transport may occur (i.e.~$I_{\tau_u} < 0$ ).

In the following we show that the destructive 
effect associated with anticorrelated regions
can be stronger than the constructive effects. 
To that end, in order to simplify the problem, 
let us consider a very simple model for the advecting velocity, $\bm{U}$, 
whose stream function is defined as:
\begin{equation}
\label{psi}
\psi(y,t) = \psi_0\,\left[ \alpha(t)\cos\left(k_0 
y\right) + \beta(t)\sin\left(k_0 y \right)\right] \;\;\; .
\end{equation}
For this $2$-d random parallel flow, 
as in Ref.~\cite{MV97},
the (random) processes $\alpha(t)$ and $\beta(t)$ are assumed Gaussian, 
independent and with  auto-correlation function given by 
\begin{equation}{\cal C}(t)= 
e^{-|t|/\tau_{\mbox{\tiny U}}}\,\cos\left(\omega t\right)
\label{MY}
\end{equation}
where $\tau_{\mbox{\tiny U}}$ is the correlation time of $\bm{U}$
and $\omega$ is the parameter controlling
the extension of anticorrelated regions.
The form of (\ref{MY}) permits us to obtain 
a finite and non-vanishing  effective diffusivity also for $D^e=0$ 
and, moreover, represents the simplest way to introduce recirculating zones 
where the scalar tends to be homogenized and trapped.\\
The following example of cellular flow defined as 
\begin{equation}
{\bm U}(x,y)=U \; (\cos(y),\cos(x))\;\;\; ,
\label{BC}
\end{equation}
the  well-known `BC' flow \cite{Ar65,He66,Doet86}
whose streamlines form a closed structure made of four cells
in each periodicity box,
shows indeed that auto-correlation functions like
(\ref{MY}) are associated to recirculation.  
In order to show this point, we have integrated the Langevin equation
by using a second-order Runge--Kutta scheme,
with the advecting velocity field 
given by eq.~(\ref{BC}) superimposed to a white-in-time noise.
The averages necessary to evaluate the auto-correlation function
$C(t)=\langle {\bm U}({\bm x}(t))\cdot {\bm U}({\bm x}(0))\rangle/U^2$,
($U^2=\langle{\bm U}\cdot {\bm U}\rangle$ is the energy of the
advecting velocity field) have been made over different realizations 
and performed by uniformly distributing $10^6$ particles 
in the four cells of the periodicity box. The obtained auto-correlation  
\begin{figure}[bt]
\begin{center}
\vspace{-1cm}
\mbox{\psfig{file=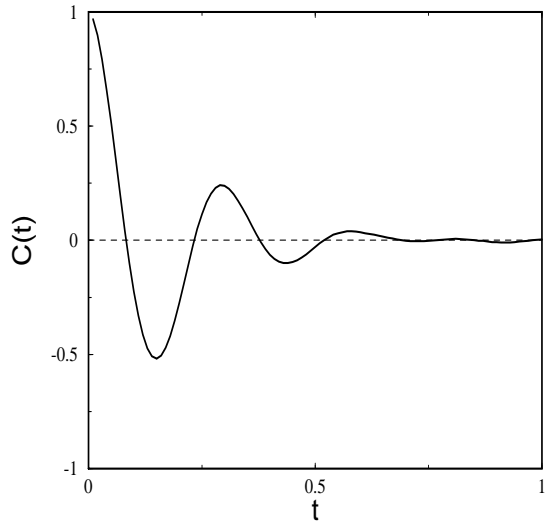,height=8cm,width=8cm}}
\vspace{-1cm}
\end{center}
\caption{The velocity auto-correlation function 
$\langle {\bm U}({\bm x}(t))\cdot {\bm U}({\bm x}(0))\rangle/U^2$
as a function of the time $t$ for the `BC' flow (\ref{BC}). 
The curve has been  obtained by integrating, 
with  a second-order Runge--Kutta scheme,
the Langevin equation for the coordinate of scalar particles 
in the presence of a white-in-time noise.}
\end{figure}
function is shown  in Fig.~1. 
The qualitative resemblance between the 
BC auto-correlation function and that given by the simple 
expression (\ref{MY}) makes meaningful the use of the latter to mimic
the statistical effect of recirculation in the diffusion process.\\
Notice that, although possible, the main 
disadvantage in studying the interference problem
by using the BC flow is that only numerical analysis seems 
to be feasible.  
This is not the case for the random 
parallel flow (\ref{psi}) with the auto-correlation (\ref{MY}), 
where the interference mechanisms can  be investigated analytically.
To that end, we use asymptotic methods since we are interested in the
dynamics of the field $\theta$ on scales of the order of $L$,
larger than $l_0 $, the typical scale of the advecting   
velocity field, ${\bm U}$.  
The ratio $l_0/L \sim O(\epsilon)$
with $\epsilon\ll 1$, the parameter controlling the
scale separation, naturally suggests to look for a perturbative
approach.  The perturbation is however singular \cite{BO78}
and multiscale techniques \cite{BLP78} are thus needed to treat 
the singularities.
For the random parallel flow, the auxiliary equation 
(a partial differential equation which needs to be solved to 
obtain the eddy-diffusivity) can be tackled analytically and a simple 
expression for the component of the effective diffusivity tensor parallel 
to the 
direction of the velocity $\bm{U}$ (the normal component being 
equal to $D^e$) can be obtained  \cite{CC98} and its expression reads: 
\begin{equation}
D^E(D^e,\tau_u)= D^e + u_0^2\int_0^{\infty} {\cal C}(t) 
\left\{e^{-k_0^2 D^e\left[t+\tau_{u}\left(e^{-\frac{t}{\tau_{u}}}-1   
\right)\right]} \right\} \; dt \;\;\; ,
\label{simple}
\end{equation}
where $u_0^2\equiv\psi_0 k_0$ and ${\cal C}(t)$ has a typical correlation time
$\tau_{\mbox{\tiny U}}$ as in (\ref{MY}).

The two opposite limits $\tau_u/\tau_{\mbox{\tiny U}}\ll 1$  and
$\tau_u/\tau_{\mbox{\tiny U}} \gg 1$ can be easily investigated 
from (\ref{simple}). About the former, it is immediately 
checked that the results obtained in Ref.~\cite{CM98} are here
reproduced by expanding at the first order in both $\tau_u$ and $D^e$
the term inside the
curly brackets in the r.h.s.~of (\ref{simple}). 
As a result, 
the interference term $I_{\tau_u}$ reads:
\begin{equation}
I_{\tau_u} = D^e k_0^2 u_0^2\tau_u\int_0^{\infty}{\cal C}(t)\;dt\;\;\; .
\end{equation}
The second limit can be easily exploited by expanding the 
same term inside the curly brackets but now for $\tau_u/\tau_{\mbox{\tiny U}} 
\gg 1$. The expression for $D^E$ corrected at the order
$1/\tau_u$ is thus obtained:
\begin{equation}
D^E = D^e + u_0^2\int_0^{\infty} {\cal C}(t) \left[ 1 - 
\frac{k_0^2D^e\tau_u}{2}\left(\frac{t}{\tau_u}\right)^2\right]\;dt\;\;\; ,
\end{equation}
from which the interference term $I$ immediately follows:
\begin{equation}
I = -\frac{u_0^2 k_0^2 D^e}{2\tau_u}\int_0^{\infty} t^2 {\cal C}(t)\;dt\;\;\; .
\label{itot}
\end{equation}
It is easy to verify from (\ref{simple}) that, for both $\tau_u=0$ and small
$D^e$ (i.e.~$D^ek_0^2\tau_{\mbox{\tiny U}}\ll 1$), the interference term
associated to $D^e$ is given by:
\begin{equation}
I_{D^e} = -D^e u_0^2 k_0^2 
\int_0^{\infty} t\, {\cal C}(t)\;dt
\label{id}
\end{equation}
and thus, from (\ref{itot}), it follows:
\begin{equation}
I_{\tau_u} = -u_0^2 k_0^2 D^e \left(\frac{1}{2\tau_u}\int_0^{\infty}
t^2 {\cal C}(t)\;dt - \int_0^{\infty}t\, {\cal C}(t)\;dt\right)\;\;\; .
\label{it}
\end{equation}
For large $\tau_u$'s, the character of the interference due to the effect of
$\tau_u$ on 
the Lagrangian trajectories is thus related to the variation of 
$I_{\tau_u}$ with $\tau_u$.\\
By inserting the expression (\ref{MY}) for ${\cal C}(t)$ 
into (\ref{it}) we obtain, after simple integrations, the 
following condition for destructive interference:
\begin{equation}
I_{\tau_u} < 0\qquad \mbox{when}\qquad \omega\tau_{\mbox{\tiny U}}>1\;\;\; .
\label{cond1}
\end{equation}
According to our previous discussions, destructive interference appears 
when the extension of anticorrelated regions (controlled by $\omega$)  
is large enough. Such phenomenon, found in the limit of large $\tau_u$'s, 
is actually present
for a wide range of values of $\tau_u$'s provided that 
$\tau_u/\tau_{\mbox{\tiny U}} \hspace{1mm}/\hspace{-4mm}\ll 1$.\\ 
To show this point, we have computed numerically the 
integral in the r.h.s.~of (\ref{simple}) from which we have
subtracted the contribution $I_{D^e} +
D^E(0,0)$ to obtain $I_{\tau_u}$. The behavior of $I_{\tau_u}$ as a function
of $\omega\tau_{\mbox{\tiny U}}$ for $\tau_u/\tau_{\mbox{\tiny U}}=0.05$
(long-dashed line), $0.5$ (dashed), $1.0$ (dotted) and $5.0$ 
(dot-dashed) is shown in Fig.~2 for $k_0^2 D^e \tau_{\mbox{\tiny U}}=0.05$
and $u_0^2=1$. 
\begin{figure}[ht]
\begin{center}
\vspace{0cm}
\mbox{\psfig{file=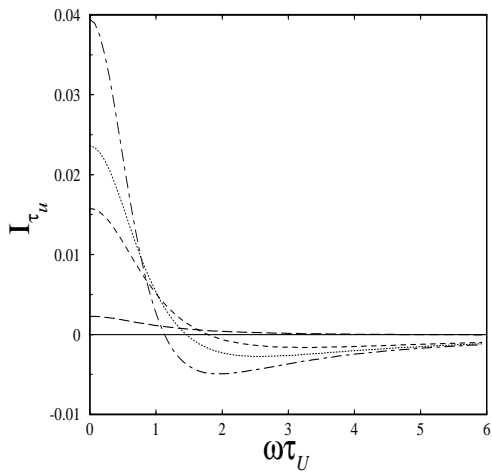,height=7cm,width=7cm}}
\vspace{-1cm}
\end{center}
\caption{The interference term $I_{\tau_u}$ as a function
of $\omega\tau_{\mbox{\tiny U}}$, for different values of
$\tau_u/\tau_{\mbox{\tiny U}}$: 0.05 (long-dashed line), 0.5 (dashed line),
 1.0 (dotted line) and 5.0 (dot-dashed). Curves have been obtained 
by numerical computation of integral \protect(\ref{simple}) with
$u_0^2=1$ and $k_0^2D^e\tau_{\mbox{\tiny U}}=0.05$.}
\end{figure}
 From this figure we observe that the smaller 
$\tau_u/\tau_{\mbox{\tiny U}}$, 
the larger the values  of $\omega \tau_{\mbox{\tiny U}}$ at which 
the crossover from constructive to destructive interference takes place.
 For all $\tau_u$'s, we can then conclude
that the condition (\ref{cond1}) should be replaced by: 
\begin{equation}
I_{\tau_u} < 0\qquad \mbox{when}\qquad \omega\tau_{\mbox{\tiny U}}>K_c
\label{congett}
\end{equation}
with $K_c$ depending on $\tau_u/\tau_{\mbox{\tiny U}}$, $K_c \geq 1$ 
(on account of (\ref{cond1})) and
$\lim_{\tau_u \to 0} K_c = \infty$ (on account of the positive character
of the interference when $\tau_u\to 0$).
Condition (\ref{congett}) is corroborated by the results of Fig.~3, where the
values,  $K_c$'s, of $\omega\tau_{\mbox{\tiny U}}$ at which the crossover
takes place, have been plotted for different values of 
$\tau_u/\tau_{\mbox{\tiny U}}$. 
\begin{figure}[ht]
\begin{center}
\vspace{0cm}
\mbox{\psfig{file=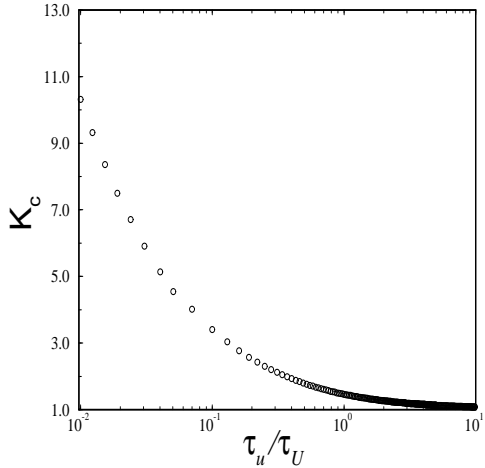,height=7cm,width=7cm}}
\vspace{-1cm}
\end{center}
\caption{Values $K_c$ of $\omega\tau_{\mbox{\tiny U}}$ corresponding
to the crossover from constructive to 
destructive interference 
(i.e.~from $I_{\tau_u} > 0 \mbox{~to~} I_{\tau_u}<0$), as 
a function of the ratio
$\tau_u/\tau_{\mbox{\tiny U}}$. Values of $K_c$ reported in figure
have been obtained 
by numerical computation of integral \protect(\ref{simple}) with
the same parameters of Fig.~2.}
\end{figure}

In conclusion, the effect on the scalar transport due to a finite
correlation time $\tau_u$ of the noise velocity
has been investigated. We have shown, both analytically 
and numerically, that for $\tau_u$ large enough two mechanisms exist 
and act in competition to determine the interference character.
The first mechanism works to enhance  the transport due to an augmentation
of coherence induced in the diffusion process.
The second mechanism maintains particles inside the zones of trapping
(where the auto-correlation of the advecting  velocity field is negative) 
for a time longer than the time relative 
to the white-noise case. Depending on the extension of 
anti-correlated regions, constructive or destructive interference take place.
For finite values of $\tau_u$, the symmetry with the scenario presented in 
Ref.~\cite{MV97} is thus perfectly restored.

\vskip 0.2cm
{\bf Acknowledgements}
We are particularly grateful to M.~Vergassola and  A.~Vulpiani
for very stimulating discussions and suggestions.
Very useful discussions with A.~Crisanti, R.~Pasmanter and E.~Zambianchi are
also acknowledged.
The work of AM has been supported by the ``Henri Poincar\'e'' fellowship 
(Centre National de la Recherche Scientifique and Conseil G\'en\'eral des
Alpes Maritimes). PC is grateful to the European Science Foundation
for the TAO exchange grant. PC has been partially  supported by INFM 
(PRA-TURBO) and by MURST (no. 9702265437).


\begin{thebibliography}{99}

\bibitem{Moffatt} H.K.~Moffatt, 
{\it Rep. Prog. Phys.}, {\bf 46}, 621 (1983).

\bibitem{BLP78}
              A.\ Bensoussan, J.-L.\ Lions and G.\ Papanicolaou,
              {\it Asymptotic Analysis for Periodic Structures}
              (North-Holland, Amsterdam, 1978)
 

\bibitem{T21} G.I.~Taylor, 
{\it Proc. Lond. Math. Soc. Ser. 2}, {\bf 20}, 196 (1921).

\bibitem{D85} R.E~Davis, {\it J. Geophys. Res.}, {\bf  90}, 4756 (1985).

\bibitem{PN89} P.M.~Poulain and P.P. Niiler, 
{\it J. Phys. Oceanogr.}, {\bf 19}, 1588 (1989).

\bibitem{Saff} P.G.~Saffman, 
{\it J. Fluid Mech.}, {\bf 8}, 273 (1960).

\bibitem{MV97} A.\ Mazzino and  M.\ Vergassola, 
{\it Europhys. Lett.},  {\bf 37}, 535 (1997).

\bibitem{CM98} P.~Castiglione and A.\ Mazzino,
{\it Europhys. Lett.},  {\bf 43}, 522 (1998).

\bibitem{Ar65} V.I.~Arnold,
               {\it C.R. Acad. Sci. Paris A}, {\bf261}, 17 (1965).

\bibitem{He66} M. H\'enon,
              {\it C. R. Acad. Sci. Paris A}, {\bf 262}, 312 (1966).


\bibitem{Doet86} T.~Dombre, U.~Frisch, J. M.~Greene, M.~H\'enon,
                 A.~Mehr and A.M.~Soward,
                 {\it J. Fluid Mech.},  {\bf 167}, 353 (1986).

\bibitem{BO78} C.M.~Bender and S.A.~Orszag, 
             {\it Advanced Mathematical Methods for Scientists and Engineers}
             (McGraw--Hill, 1978).

\bibitem{CC98}
P.~Castiglione and A.~Crisanti,
``Dispersion of passive tracers for a Langevin equation 
with non delta-correlated noise'',
submitted to {\it Phys. Rev. E}, (1998).


\end{thebibliography}
\end{document}